%% file: paper.tex
\documentclass[preprint2,numberedappendix,iop]{emulateapj-rtx4}
\usepackage{graphicx,graphics,amsmath}
\usepackage{natbib}
\usepackage{bm,url}
\usepackage{times}
\usepackage{amssymb}
\usepackage{color}
\graphicspath{{./fig/}{./png/}}
\input aa

\shorttitle{Small-scale field and dynamo cycle}
\shortauthors{Karak \& Brandenburg}

\begin{document}
\title{Is the small-scale magnetic field correlated with the dynamo cycle?}

\author{Bidya Binay Karak$^{1,2}$ and Axel Brandenburg$^{1,3,4,5}$}
\affil{$^1$Nordita, KTH Royal Institute of Technology and Stockholm University, Roslagstullsbacken 23, SE-10691 Stockholm, Sweden\\
$^2$Max-Planck-Institut f\"ur Sonnensystemforschung, Justus-von-Liebig-Weg 3, D-37077 G\"ottingen, Germany\\
$^3$Department of Astronomy, AlbaNova University Center, Stockholm University, SE-10691 Stockholm, Sweden\\
$^4$Department of Astrophysical and Planetary Sciences, University of Colorado, Boulder, CO 80303, USA\\
$^5$Laboratory for Atmospheric and Space Physics, University of Colorado, Boulder, CO 80303, USA
}

\begin{abstract}
The small-scale magnetic field is ubiquitous at the solar surface---even
at high latitudes.
From observations we know that this field is uncorrelated (or perhaps
even weakly anticorrelated) with the global sunspot cycle.
Our aim is to explore the origin, and particularly the cycle dependence 
of such a phenomenon using three-dimensional dynamo simulations.
We adopt a simple model of a turbulent dynamo in a shearing box driven by helically
forced turbulence. Depending on the dynamo parameters, large-scale (global) 
and small-scale (local) dynamos can be excited independently in this model.
Based on simulations in different parameter regimes, we find
that, when only the large-scale dynamo is operating in the system,
the small-scale magnetic field generated through shredding and tangling
of the large-scale magnetic field is positively correlated with the global
magnetic cycle.
However, when both dynamos are operating, the small-scale field is produced
from both the small-scale dynamo and the tangling of the large-scale field.
In this situation, when the large-scale field is weaker than the 
equipartition value of the turbulence, the small-scale field is almost uncorrelated with the
large-scale magnetic cycle.
On the other hand, when the large-scale field is stronger than 
the equipartition value, we observe an anticorrelation between
the small-scale field and the large-scale magnetic cycle.
This anticorrelation can be interpreted as a suppression of the
small-scale dynamo.
Based on our studies we conclude that the observed small-scale
magnetic field in the Sun is generated by the combined mechanisms of
small-scale dynamo and tangling of the large-scale field.
\end{abstract}

   \keywords{Magnetohydrodynamics --
                turbulence --
                dynamos --
                Sun: magnetic fields --
%                Sun: Solar cycle}
%BBK: Changed as this is not allowed in ApJ.
                sunspots}
  \email{bbkarak@nordita.org and bidyakarak@gmail.com}
   \maketitle
%____________________________________________________________

\section{Introduction}

Sunspots and the solar cycle are associated with the large-scale magnetic field 
of the Sun. This magnetic field is believed to be produced by a large-scale dynamo, 
{\it globally} operating in the solar convection zone. 
The $\alpha$ effect (originating either from helical flow and/or the Babcock-Leighton process) 
and the $\Omega$ effect (shearing of the magnetic field by differential rotation) are
believed to be the cause of this dynamo in the Sun \cite[see reviews by][]{C14,K14}.
However, there is another type of magnetic field that exists at small scales
and even in the quiet phase of the Sun.
This field is intermittent and fluctuating and 
exists in mixed polarity even at the resolution limits of present day instruments. 
This field is known as small-scale, fluctuating, turbulent,
or internetwork field---as it exists in the internetwork regions. 
Although the existence of this field has been known for a long time
\citep[e.g.,][]{HS72, FS72}, 
the detailed understanding of its nature and origin has remained elusive.
The main reason for
this is that the spatial characteristic scale of this field is shorter than the 
resolution limits of present day instruments and the weak 
polarization signal makes the measurement of this quiet-Sun magnetic field 
difficult \cite[e.g.,][]{de09}. 
Interestingly, even the internetwork field---the weakest component 
of solar magnetism---having an unsigned flux of $\approx 10^{15}-10^{18}\Mx$,
contributes about $10^{26}$~Mx of magnetic flux 
to the solar surface per day, which is at least four orders of magnitude 
more than that from the bipolar active regions
(which contain an unsigned flux of $\approx 5\times 10^{20} - 5\times 10^{22}\Mx$)
during solar maximum \citep{SH94,ZWJ13}.
In comparison, both the network regions with a flux of $\approx 10^{17}-10^{19}\Mx$
and ephemeral regions with $10^{18}-5\times10^{20}\Mx$ contribute roughly two
orders of magnitude more flux than active regions \citep{HST03,Go14}.
Therefore, the small-scale magnetic field is potentially important in understanding
many unsolved problems in solar physics, particularly solar atmospheric heating \citep[e.g.,][]{Sch98}.

Since \citet{Ba50} it is known that a sufficiently random 
three-dimensional velocity field can amplify a
seed magnetic field via repeated random stretching, bending and folding---without requiring 
net helicity.
This is the essence of the small-scale or local dynamo.
The correct mathematical description of such dynamos
was developed later by \cite{Kaz68}; see also \citet{CG95}.
\citet{PS93} were the
first to realize the necessity of a small-scale dynamo to explain the magnetic flux 
budget in the photosphere \citep[also see][]{DDR93}.
However, one may put forward alternative arguments in which this quiet-Sun 
magnetic field is the result of a global large-scale dynamo: 
the shredding of large-scale magnetic 
field can produce small-scale magnetic field by cascading magnetic energy to small-scales
and the decay of active regions may give rise to small-scale magnetic field \citep{STB87,deW05,S12}.
However, none of these arguments may necessarily be correct because
the small-scale magnetic field does not show solar cycle dependence,
as it does not have significant
correlation with the large-scale global magnetic cycle, and 
it does not have much latitudinal variation
\citep[e.g.,][]{HST03,Sa03,L08,Li11,BLS13,JW15a}. 
Furthermore, using the MDI data set of the entire solar cycle 23, \citet{Jin11}
have shown that the cyclic variation of both number and total flux of
small-scale magnetic elements in the range 
$\sim (3$--$30) \times 10^{18}\Mx$
are anticorrelated with the sunspot number.
Subsequently, \citet{JW15b} have studied the internetwork
magnetic field within the flux range of about $10^{16}$--$10^{18}\Mx$ and
find no correlation with the solar cycle. All these observational results support the
small-scale dynamo scenario as the source of small-scale field in the Sun.
We note that
\citet{SKA15} argued in favor of the operation of a small-scale dynamo
in the Sun to explain the observational findings that the
number of anti-Hale sunspot groups becomes maximal during solar minimum and
that these groups do not show the latitudinal dependence as Hale sunspots\footnote{
One may argue that the anti-Hale sunspot groups could be
formed through the negative effective magnetic pressure instability
\citep{BKKMR11,BKR13,KBKMR13} because,
for this to work, we only need a moderately weak magnetic field, which could
be provided by the small-scale dynamo operating continuously in the Sun.
The anti-Hale groups would then represent just the tail of a wide
distribution of tilt angles \citep{McClintock+Norton+Li14}.
Within the rising flux tube picture, anti-Hale groups near the equator
could alternatively be explained as tubes buffeted from the other hemisphere.
}.

The complex nature of the stratified convection 
zone with very low values of molecular viscosity, diffusivity, 
and magnetic Prandtl number ($\Pm$) make the theoretical modeling of
the small-scale field challenging at present. 
Furthermore, at small $\Pm$, the critical magnetic 
Reynolds number, $\Rm^C$ needed to excite the small-scale dynamo increases at first linearly
with deceasing $\Pm$ \citep{RK97,BC04,Sch05} until $\Rm^C$
saturates for $\Pm \ll 1$ \citep{Isk07, Br11},
which makes the small-scale dynamo at the solar value of $\Pm \sim 10^{-5}$ possible.
Beside these difficulties several authors \citep[e.g.,][]{C99,EC01}
tried to model the small-scale dynamo in a local Cartesian box.
Although they found in some cases
small-scale dynamo action, these simulations are highly idealized and far 
from the real Sun. Subsequently, \cite{V05} \citep[see also][]{VS07,PCS10,Rem14} have gone one step further
to make these models more realistic by including radiation and ionization. 
However, these simulations are local in the sense that they cannot study the effects of
the global dynamo-generated large-scale field on the small-scale field.
Very recently, \citet{HRY14,HRY15} have studied small-scale dynamo action in the Sun
using high-resolution MHD simulations.
Again, however, these simulations did not consider the large-scale dynamo.

The purpose of the present study is to explore the effects of a global dynamo-generated 
cyclic large-scale magnetic field on the small-scale field. 
To this end, we perform a set of simulations of a turbulent cyclic dynamo
that produces a migratory magnetic field similar to that of the Sun.
To capture the essential features of an $\alpha\Omega$ dynamo,
in particular its migratory properties, we employ helically forced
turbulence together with imposed uniform shear across the domain.
These two ingredients provide a prototype of an $\alpha\Omega$ dynamo,
as it is often modelled in mean-field theory.
Earlier turbulence simulations of this type have confirmed the presence
of cyclic solutions with dynamo waves whose direction of propagation
depends on the sign and magnitude of shear and helicity \citep{KB09,KKB15}.

We do not make an attempt to model the real Sun, whose
underlying dynamo mechanism is still not well understood \citep{Br05,C14}.
It is not even completely clear whether the solar dynamo is driven
by convective motions or by magnetic instabilities \citep{Br98,TCB11},
and whether meridional circulation plays an important role for causing
the equatorward migration of the Sun's magnetic field \citep{CSD95,HKC14}.
Furthermore, we are facing even more fundamental challenges, 
for example the question of catastrophic
quenching, the dependence of mean field strength on magnetic Reynolds
number, and the morphology of the field in the form of tubes on the
one hand, and the question of magnetic buoyancy, flux tube storage,
and sunspot formation on the other. 
Significant progress has been
made by considering these aspects separately, since in models with all
effects included it is difficult to associate particular features to
given changes in the model. Thus, in our model, we focus on the aspect
of large-scale and small-scale dynamos only, i.e., we do not consider the
question of magnetic buoyancy, flux tube storage, and sunspot formation.
Consequently, we consider a model without gravity and convection
and consider turbulence to be driven by a stochastic forcing.
Moreover, our simple model then avoids the vastly different time and length 
scales of global and local dynamos in realistic settings.
Capturing them within a single model remains a challenging task
in global convection simulations \citep{V05,KMCWB13,Kar15,HRY15}.
In the present paper, our aim is to address the basic question whether and how the
small-scale magnetic field is correlated with the
dynamo cycle in a simple realization of a turbulent dynamo that combines
the physics of a small-scale dynamo with that of an $\alpha\Omega$ dynamo.

\section{Model}

In this work we do not consider convection.
Therefore, gravitational stratification and thermodynamics are neglected.
Instead, our model consists of an isothermal, compressible fluid where the
pressure is given by $p=\cs^2 \rho$, with $\cs$ being the isothermal (constant) sound speed
and $\rho$ the density.
We sustain turbulence by supplying energy to the system through a
stochastic forcing function in the momentum equation.
As we do not include the stratification and global rotation of the Sun, we make this forcing
helical to produce an
$\alpha$ effect in our model. Furthermore, to mimic the large-scale
shear of the Sun and to produce an $\Omega$ effect, we impose a uniform
shear flow in our model.
We solve the hydromagnetic equations in a cubic domain of size $(2\pi)^3$.
To connect with spherical geometry, we can imagine
our simulation box being located in the northern hemisphere of the Sun
such that the $x$ axis corresponds to the radially outward direction, 
the $y$ axis to the toroidal direction, and
the $z$ axis to the direction of increasing latitude;
see, e.g., \cite{KMCWB13}, who used the same correspondence in their Section~3.6.
Within this framework, the momentum, continuity, and induction equations become
\begin{equation}
\frac{D \UU}{D t} = \ff-S U_x \hat y
-c_{\rm s}^2\nab\ln\rho
+ \frac{1}{\rho} \left[\JJ \times \BB +
\nab\!\cdot(\!2\rho\nu\SSSS)\right],\;
\label{eqmom}
\end{equation}
\begin{equation}
\frac{D\ln\rho}{D t} =-\nab\cdot\UU,
\end{equation}
\begin{equation}
\frac{\partial \AAA}{\partial t} + \UU^{(S)} \cdot \nab \AAA 
= - S A_y \hat x + \UU\times\BB +  \eta \nab^2 \AAA,
\end{equation}
where
$D/Dt = \partial/\partial t + (\UU  + \UU^{(S)})\cdot \bm\nab$
is the advective time derivative, 
$\UU^{(S)}=(0,Sx,0)$ is the imposed uniform large-scale shear flow with
$S$ being the constant shear rate, $\BoldVec{f}$ is the forcing function that we describe later, 
$\nu$ is the constant kinematic viscosity, and
$\SSSS$ is the traceless rate of strain tensor given by
${\sf S}_{ij} = \onehalf (U_{i,j}+U_{j,i}) - \onethird \delta_{ij} \DIV \bm{U}$
(the commas denote partial differentiation with respect to the
coordinate $j$ or $i$).
The contribution of $\UU^{(S)}$ to the viscous stress is small compared with
that of $\UU$ and is neglected.
We note that the induction equation is written in terms of the magnetic vector potential
$\AAA$, such that $\nab\times {\AAA}$ = ${\BB}$ is the magnetic field. 
Furthermore, $\JJ = \mu_0^{-1} \nab\times\BB$ is the current density, and
$\eta$ is the constant microscopic diffusivity

The forcing function $\BoldVec{f} = \BoldVec{f}(\xx,t)$ in \Eq{eqmom} is helical
and random in time.
It is defined as
$\ff(\xx,t)={\rm Re}\{N\ff_{\kk(t)}\exp[\ii\kk(t)\cdot\xx+\ii\phi(t)]\},$
where $\xx$ is the position vector.
At each timestep, the wavevector $\kk(t)$ randomly takes any value from
all possible wavevectors in a certain range around a given forcing wavenumber
$k_{\rm f}$.
The phase $-\pi<\phi(t)\le\pi$ also changes randomly at every timestep.
On dimensional grounds, we choose $N=f_0 c_{\rm s}(|\kk|c_{\rm s}/\delta t)^{1/2}$,
where $f_0$ is a non-dimensional forcing amplitude.
The transverse helical waves are produced via Fourier amplitudes \citep{Hau04}
\begin{equation}
\ff_{\kk}=\RRRR\cdot\ff_{\kk}^{\rm(nohel)}\quad\mbox{with}\quad
{\sf R}_{ij}={\delta_{ij}-\ii\sigma\epsilon_{ijk}\hat{k}_k
\over\sqrt{1+\sigma^2}},
\label{eq: forcing}
\end{equation}
where $\sigma$ is a measure of the helicity of the forcing; for positive maximum helicity,
$\sigma=1$, 
$\ff_{\kk}^{\rm(nohel)}=
\left(\kk\times\eee\right)/\sqrt{\kk^2-(\kk\cdot\eee)^2}, \nonumber$
where $\eee$ is an arbitrary unit vector
not aligned with $\kk$. Note that $|\ff_{\kk}|^2=1$ and
$\ff_{\kk}\cdot(\ii\kk\times\ff_{\kk})^*=2\sigma k/(1+\sigma^2)$.

We solve the above equations with the {\sc Pencil Code}\footnote{
\url{http://github.com/pencil-code}} in $-\pi < x,y,z<\pi$
with boundary conditions that are shearing--periodic
in the $x$ direction and periodic in the $y$ and $z$ directions.
The numerical time step $\Delta t$ is computed as the Courant time step.
As initial conditions we take $\uu=\ln\rho=0$
and $\AAA$ consists of small-scale low amplitude ($10^{-4}$)
Gaussian noise.
In our nondimensional unit system, we have $\cs=\rho_0=\mu_0=1$, where
$\rho_0 = \bra{\rho}$ is the volume-averaged density, which is constant
in time because of mass conservation.

To interpret our results in terms of a mean-field picture,
we define the mean-field dynamo number $D = C_\alpha C_\Omega$ with 
\EQ
C_\alpha=\alpha_0/\etaTz k_1 \quad
{\rm and} \quad
C_\Omega=|S|/\etaTz k_1^2,
\label{eq:DN}
\EN
where $\etaTz=\etatz+\eta$ is the theoretically expected total
(turbulent plus microphysical) magnetic diffusivity, 
$\etatz=\tau \langle \uu^2 \rangle/3$,
$\alpha_0=-\tau \langle \boldsymbol \omega \cdot \uu \rangle/3$,
$\tau=(\urms\kf)^{-1}$,
and $\kf$ is the mean forcing wavenumber \citep{BB02}.
The other usual diagnostic parameters, $\Rm$, $\Rey$ (fluid Reynolds number), and $\Pm$ are defined as
\EQ
\Rm=\urms/\eta\kf,\quad
\Rey=\urms/\nu\kf, \quad
\Pm=\nu/\eta, \quad
\EN
where $\urms =\langle \uu^2 \rangle^{1/2}$ is the rms value of the velocity
in the statistically stationary state with
$\langle\cdot\rangle$ denoting the average over the whole domain and over time, and
$\Beq = \urms$ is the volume-averaged equipartition field
(we recall that in our units, $\mu_0\bra{\rho}=1$).
Furthermore, the large-scale (mean) field will be defined as a horizontal average
and therefore the small-scale field will be defined as 
the residual of the total and the large-scale quantities viz,
\EQ
\overline{\bb^2}=\overline{\BB^2}-{\overline{\BB}}^2.
\label{def:ssf}
\EN
Here again overbars denote $xy$-averages.

\begin{table*}[t!]
\centering
\caption[]{Summary of the runs.}
      \label{tab:runs}
      \vspace{-0.5cm}
     $$
         \begin{array}{p{0.05\linewidth}p{0.03\linewidth}rlcllccccccccrrcl}
           \hline
           \hline

& & \multicolumn{5}{c}{\rm Input}& & & & & & & {\rm Output} \\
\cline{1-7}
\cline{9-19}
           \noalign{\smallskip}
\rm {Regime} & \rm Run & \rm {grid} &~\kf & \nu~~&~\Pm\;\;&|S|&& \urms/\cs&\Rm &\tilde{\meanB}_{\rm x}&\tilde{\meanB}_{\rm y} & \tilde{\meanB}& \tilde{\mean b}& C_\alpha & C_\Omega & D~ &{{\rm Corr}(\meanB_y^2,\overline{\bb^2})} & {\rm Dynamo?}\\ \hline

      I  & R1  & 48^3 & 5.1 &10^{-2}& 1.25&0.05&& 0.023 & 0.55 & 0.000 & 0.00 & 0.00 & 0.000 & 0.78 &  5.3 & 4.1 & -- &\rm No~dynamo \\%Run folder 48g
\hline
{\bf II} & R2a & 48^3 & 5.1 &10^{-2}& 1.25 &0.1&& 0.022 & 0.55 & 0.165 & 1.40 & 1.41 & 0.274 & 0.75 & 10.6 & 7.9 &~0.77 & \rm LSD\\ %48f
{\bf II} & R2b & 96^3 & 3.1 &5\times10^{-4}& 0.42&0.05&& 0.066 & 17.7 & 0.246 & 1.34 & 1.38 & 0.662 & 1.38 &  6.1 &  8.4 &~0.42 & \rm LSD \\%pm0p5_96a
{\bf II} & R2c & 96^3 & 3.1 &5\times10^{-4}& 0.50 &0.04&& 0.063 & 20.1 & 0.298 & 1.50 & 1.54 & 0.711 & 1.58 &  5.2 &  8.2 &~0.48 & \rm LSD \\%pm0p5_96b
{\bf II} & R2d & 48^3 & 5.1 &10^{-2}& 1.25 &0.15&& 0.022 & 0.54 & 0.202 & 2.57 & 2.61 & 0.418 & 0.71 & 15.9 & 11.4 &~0.91 & \rm LSD \\%48fs0p15
{\bf II} & R2e & 48^3 & 5.1 &10^{-2}& 1.25 &0.2&& 0.039 & 0.97 & 0.294 & 4.56 & 4.57 & 0.671 & 1.14 & 18.9 & 21.6 &~0.76 & \rm LSD \\%48fs0p2
{\bf II} & R2f & 48^3 & 5.1 &10^{-2}& 1.25 &0.25&& 0.039 & 0.97 & 0.271 & 5.17 & 5.18 & 0.699 & 1.11 & 23.6 & 26.2 &~0.70 & \rm LSD \\%48fs0p25
{\bf II} & R2g & 48^3 & 5.1 &10^{-2}& 5.00 &0.08&& 0.021 & 2.06 & 0.308 & 7.17 & 7.18 & 0.862 & 2.00 & 23.7 & 47.4 &~0.62 & \rm LSD \\%48bb
{\bf II} & R2h & 48^3 & 5.1 &10^{-2}& 5.00 &0.1&& 0.021 & 2.06 & 0.282 & 8.06 & 8.07 & 0.884 & 1.98 & 29.7 & 58.8 &~0.50 & \rm LSD \\%48b
\hline

     III & R3  &144^3& 3.1 &10^{-3}& 5.0 &0.0&& 0.054 & 86.1 & 0.076 & 0.11 & 0.15 & 0.526 & 0.00 &  0.0 & 0.0 & -- & \rm SSD \\
     III & R3$^\prime$  &288^3& 3.1 &10^{-3}& 5.0 &0.0&& 0.054 & 86.1 & 0.076 & 0.11 & 0.15 & 0.528 & 0.00 &  0.0 & 0.0 & -- & \rm SSD \\

\hline

%---Regime IV-----
{\bf IV} & R4a &144^3& 3.1 &10^{-3}& 5.0 &0.05&& 0.057 & 90.7 & 0.216 & 1.83 & 1.88 & 0.882 & 2.07 &  8.0 & 16.6 & -0.57 & \rm{SSD+LSD}\\ %Run folder: karak/forced/shear/144c
{\bf IV} & R4a$^\prime$ &288^3& 3.1 &10^{-3}& 5.0 &0.05&& 0.057 & 90.6 & 0.235 & 1.88 & 1.90 & 0.937 & 2.14 &  8.0 & 17.2 & -0.56 & \rm{SSD+LSD}\\ %Run folder: karak/forced/shear/144c_288new
{\bf IV} & R4b &144^3& 3.1 &10^{-3}& 5.0 &0.1&& 0.065 & 104  & 0.155 & 1.42 & 1.45 & 0.835 & 1.41 & 14.1 & 19.8 & -0.53 & \rm{SSD+LSD}\\ %forced/shear/144_axel128a/
{\bf IV} & R4c &144^3& 3.1 &4\times10^{-4}& 5.0 &0.05&& 0.059 & 234  & 0.226 & 1.72 & 1.74 & 0.983 & 2.38 & 7.9 & 18.8 & -0.69& \rm{SSD+LSD}\\%144c2
{\bf IV} & R4d &144^3& 3.1 &2\times10^{-4}& 1.0 &0.05&& 0.069 & 110  & 0.231 & 1.65 & 1.68 & 0.831 & 1.87 & 6.7 & 12.5 & -0.65& \rm{SSD+LSD}\\%pm1_144c
{\bf IV} & R4e &144^3& 3.1 &10^{-4}& 0.5 &0.05&& 0.069 & 110  & 0.230 & 1.56 & 1.58 & 0.828 & 1.92 & 6.6 & 12.7 & -0.63& \rm{SSD+LSD}\\%pm0p5_144c
{\bf IV} & R4f &144^3& 3.1 &4\times10^{-4}& 5.0 &0.05&& 0.056 & 225 & 0.211 & 1.44 & 1.47 & 1.005 & 1.99 & 8.23 & 16.4 & -0.62& \rm{SSD+LSD}\\%144c2_sigp5_sp005
{\bf IV} & R4g &144^3& 3.1 &10^{-3}& 5.0 &0.3&& 0.170 & 271  & 0.059 & 0.53 & 0.53 & 0.486 & 0.14 & 16.4 &  2.3 & ~~~0.01& \rm{SSD+LSD}\\%Run:144b
{\bf IV} & R4h &144^3& 3.1 &10^{-3}& 5.0 &0.2&& 0.162 & 258  & 0.062 & 0.39 & 0.40 & 0.456 & 0.16 & 11.5 &  1.9 & ~~~0.14& \rm{SSD+LSD}\\%Run:144a
\hline
         \end{array}
     $$
\tablecomments{
The $\kf$ is in unit of $k_1 = 2\pi/L_x=1$.
Except for Run R3 for which $\sigma = 0$, for all runs we have $\sigma = 1$.
For Runs~R2e and R2f we have $f_0 = 0.02$ while for all other runs $f_0 = 0.01$.
The $\tilde{\meanB}$ is the temporal mean in the statistically
stationary state of the large-scale magnetic field over the whole domain,
$\mean{B}_{\rm rms}= \langle \langle B_x \rangle _y^2 + \langle B_y \rangle _y^2 + \langle B_z \rangle _y^2 \rangle _{xzt}^{1/2}$, 
normalized by $B_{\rm eq}$. 
Similarly, $\tilde{\mean{B}}_i=\langle \mean{B}_i^2 \rangle ^{1/2}/\Beq$
for $i=x$ and $y$, and
  $\tilde{\mean{b}}=\langle \overline{\bb^2} \rangle ^{1/2}/\Beq$.
Here, Corr($\meanB_y^2,\overline{\bb^2}$) is the linear correlation coefficient $r$ 
between $\meanB_y^2$ and $\overline{\bb^2}$.
SSD and LSD stand for small-scale and large-scale dynamos.
Bold fonts in the left column indicate regimes of our primary interest.
}\end{table*}

\section{Results}
\begin{figure}[t]\centering
\includegraphics[width=0.7\columnwidth]{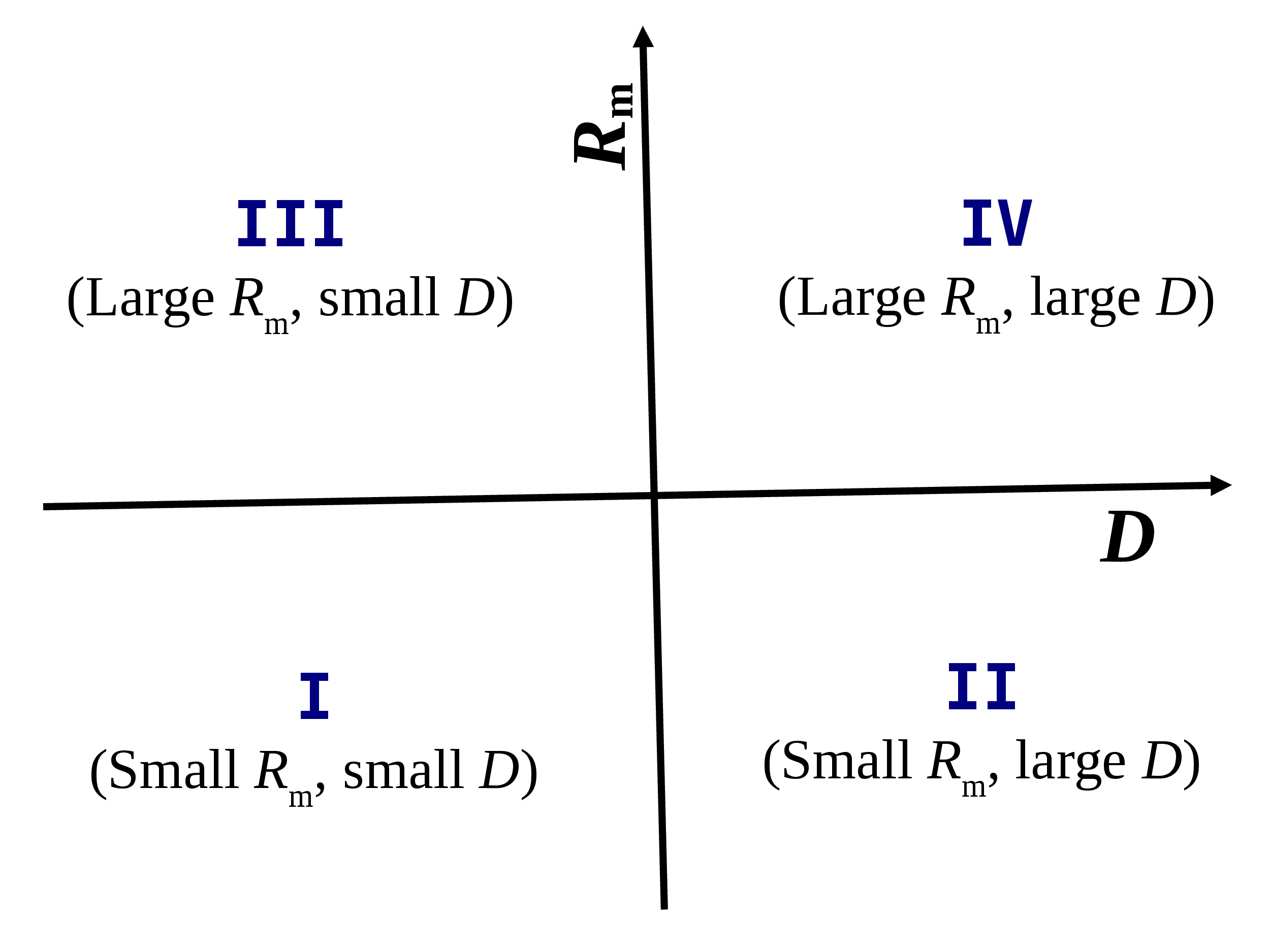}\caption{
Different regimes in magnetic Reynolds and dynamo numbers space 
to be studied in the present work.
}\label{fig:regimes}\end{figure}

We perform a set of simulations in four different regimes in the parameter 
space $(\Rm,D)$. 
We define the Regimes~I, II, III, and IV as marked in \Fig{fig:regimes}.

\subsection{Small $\Rm$ and small $D$: Regime I}
In this regime, as both the Reynolds and the dynamo numbers are small, neither 
the small-scale dynamo nor the large-scale dynamo are possible.
Therefore, any seed magnetic field decays with time.
In \Tab{tab:runs}, we show a typical run: R1 at $\nu =0.01$, $\eta =0.008$,
$\sigma = 1$ (fully helical forcing of positive helicity), and $S = - 0.05$,
so $\Rm=0.55$ and $D=4.1$, resulting in a slow decay at a rate
$0.03 \urms k_1$.

\subsection{Small $\Rm$ and large $D$: Regime II}
\label{sec:regII}

\begin{figure}[t]\centering
\includegraphics[width=\columnwidth]{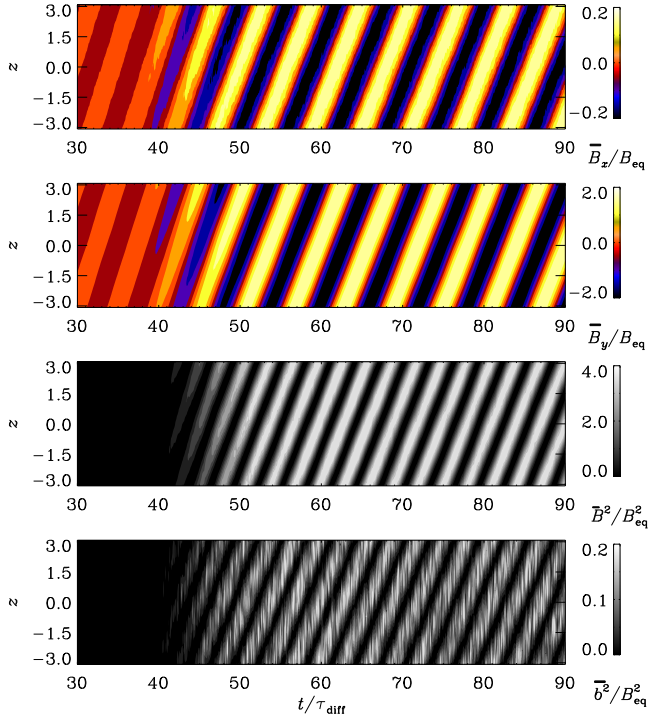}
\caption{Results from Run~R2a in Regime~II: 
Butterfly diagrams of $\meanB_x$, $\meanB_y$, $\overline{\BB}^2$, 
and $\overline{\bb^2}$ as functions of $z$ and $t$,
normalized by the diffusive time scale $\tau_{\rm diff}=(k_1^2\eta)^{-1}$ 
($k_1$ being the smallest possible wavenumber in the box).
The field is always normalized by $\Beq$.
The initial phase for $t/\tau_{\rm diff} < 30 $ is not shown, because the magnetic field
is still too weak.
}\label{fig:R3}\end{figure}

Next, we consider a set of parameters in Regime~II.
Run~R2a is a typical simulation in this regime which is summarized in \Tab{tab:runs}.
In the two top panels of \Fig{fig:R3}, we show the $x$- and $y$-components of
the mean field, $\meanB_x$ and $\meanB_y$.
We see that the large-scale field is oscillatory with dynamo waves
propagating in the positive $z$ direction.
This is indeed expected because we force the flow with positive helicity,
which results in a negative $\alpha$ effect, and since shear is also negative,
the dynamo number is positive, resulting in dynamo wave propagation
in the positive $z$ direction \citep[see, e.g.,][]{BS05}.
The toroidal field ($\meanB_y$, streamwise direction) is about 10 times
stronger than the poloidal field ($\meanB_x$, cross-stream direction);
see \Fig{fig:R3}.
This is because of the fact that the $\Omega$ effect, as measured by $C_\Omega$,
is much larger than the $\alpha$ effect, as measured by $C_\alpha$; see \Tab{tab:runs}.

\begin{figure}[t]\centering
\includegraphics[width=\columnwidth]{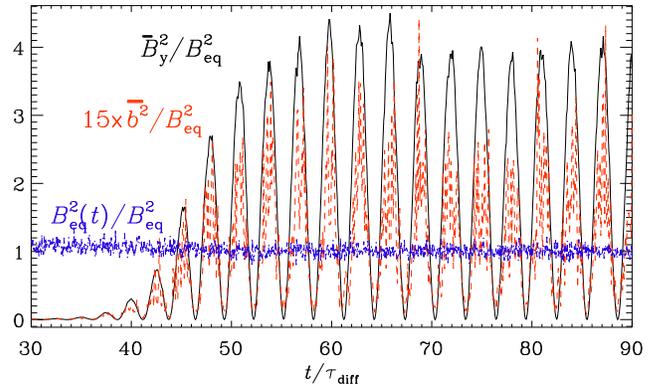}
\caption{Results from Run R2a: Time series of $\overline{B}_y^2/\Beq^2$ 
(solid black line), $\overline{\bb^2}/\Beq^2$ (red dashed), and $\Beq^2(t)/\Beq^2$ (blue dash-dotted). 
Note that $\overline{\bb^2}/\Beq^2$ is rescaled by a factor of 15.
}\label{fig:R3ts}\end{figure}

The next panel of \Fig{fig:R3} shows the energy density of the large-scale
field, $\overline{\BB}^2$, and the last panel that of the small-scale
field, $\overline{\bb^2}$, as computed from \Eq{def:ssf}.
An interesting feature to note is that the small-scale field
is spatially and temporally correlated with the large-scale one.
We also see that $\overline{\bb^2}$ is much smaller than $\overline{\BB}^2$. 
The time series of $\meanB_y^2$ and $\overline{\bb^2}$ shown in \Fig{fig:R3ts} show a tight
correlation.
Moreover, the small-scale field becomes significant only
when the large-scale field does.
In the early phase, this small-scale field grows 
linearly in time, contrary to the exponential growth 
when the field is generated through the small-scale dynamo.

\begin{figure}[t]\centering
\includegraphics[width=\columnwidth]{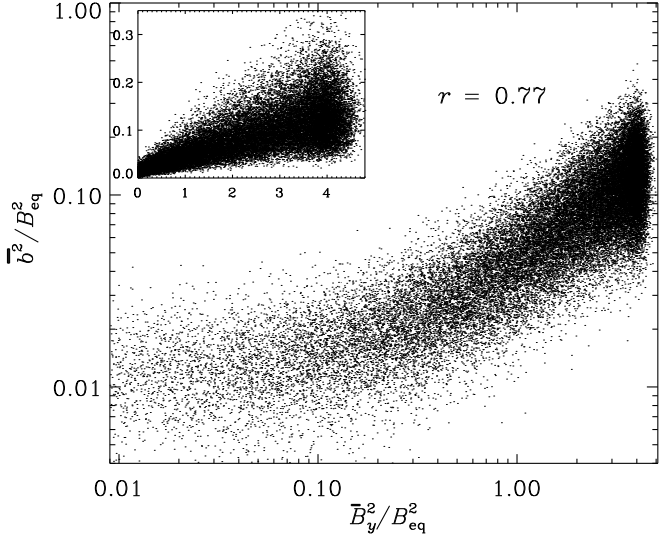}
\caption{Results from R2a: Scatter plot between $\meanB_y^2/\Beq^2$
and $\overline{\bb^2}/\Beq^2$ at $t/\tau_{\rm diff} > 40$
(nonlinear phase) for all $z$.
The inset displays the same variation, but in linear scale.
The linear correlation coefficient $r$ between these two energies is $0.77$.
}\label{fig:R3sct}\end{figure}

A scatter plot between $\meanB_y^2$ and $\overline{\bb^2}$ for all $z$ values shows
a clear positive correlation with $\overline{\bb^2} \approx q \meanB_y^2$;
see \Fig{fig:R3sct}.
It is interesting to compare this with results from the second order
correlation approximation, which yields $q\approx\Rm^2$ when $\Rm\ll1$
and the field is sufficiently weak; see Sect.~3.11 of \cite{KR80}.
However, here we find $q\sim 0.03$, which is about one order of magnitude
smaller than $\Rm^2$ ($\sim0.3$).
One reason may well be that here the large-scale field is not
imposed, as assumed in the analytic theory, but is the result of
the large-scale dynamo.

Our results demonstrate that in the present regime of small $\Rm$, the
small-scale field is the result of tangling of the large-scale field
and it is not due to the small-scale dynamo.
We recall that the small-scale field developed in the global convection simulation
of \citet{RCGBS11} in spherical geometry also shows a positive correlation
with the cyclic large-scale magnetic field.
This suggests that the small-scale
field in their simulation is also the result of tangling of the large-scale field rather
than the local dynamo.

\begin{figure}[t]\centering
\includegraphics[width=\columnwidth]{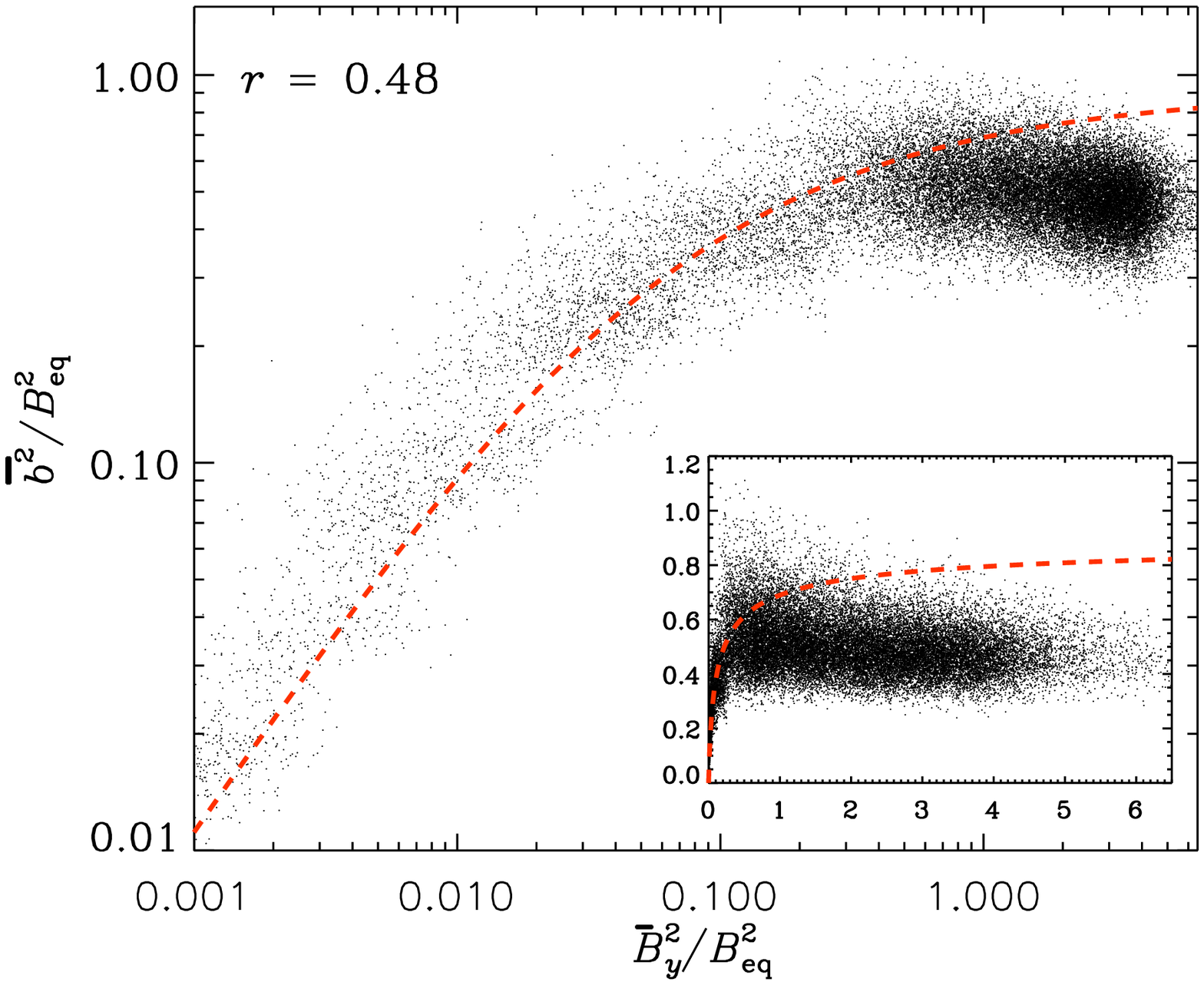}
\caption{Same as \Fig{fig:R3sct}, but from Run R2c.
The red dashed line shows
\Eq{RKfit}---the theoretical variation.
}\label{fig:tang_sat}\end{figure}

We have performed additional simulations in this regime by changing the parameters ($\kf$, $\nu$, $\eta$, or $S$); 
see Runs~R2b--R2h in \Tab{tab:runs}. 
We notice that in Runs R2a, R2b and R2c,
the large-scale magnetic field ($\tilde{\meanB}$) has not changed much 
as $D$ is similar in these runs,
but $\Rm$ and the small-scale field ($\tilde{\mean{b}}$) have changed largely. 
The increase of $\Rm$ in these runs increases the tangling. 
Therefore, the small-scale magnetic field increases with increasing $\Rm$. 
Another interesting result we notice is that, unlike in the previous case at small $\Rm$ (\Fig{fig:R3sct}),
tangling does not increase all the way with the large-scale field.
To demonstrate this we show a scatter plot between $\meanB_y^2$
and $\overline{\bb^2}$ for Run R2c---the case with the strongest $\Rm$ in \Fig{fig:tang_sat}.
Here we clearly observe the saturation of small-scale field. We also
notice that when $\meanB_y^2$ just exceeds $\Beq^2$ there is a small decrease of 
$\overline{\bb^2}$---this is caused by the fact that, at this field strength, the 
flow is significantly suppressed which reduces the tangling. Then for $\meanB_y^2 > \Beq^2$,
$\overline{\bb^2}$ remains roughly constant.

The magnetic fluctuations produced in our simulations can be compared with
the analytical theory of \citet{RK07} who have studied the magnetic fluctuations
generated by tangling of the mean magnetic field through velocity fluctuations.
Although in our simulations the values of $\Rm$ are rather small, 
we may try to compare with their analytical expression:
\begin{equation}
\overline{\bb^2} = \overline{\bb^2}^{(0)} + c_1 \left(\Beqtwoz-\overline{\bb^2}^{(0)}\right) f(c_2 \beta),
\label{RKfit}
\end{equation}
with coefficients $c_1 = 1/12$, $c_2=4$, and $\beta = \meanB_y/\Beq$, as well as
the quenching function $f(\beta)=6 - 3A_1^{(0)}(\beta) - A_2^{(0)}(\beta)$
where $A_1^{(0)}(\beta)$ and $A_2^{(0)}(\beta)$ are given
by
\begin{eqnarray}
A_1^{(0)}(\beta) = \frac{1}{5}\left[2+2\frac{\tan^{-1}\beta}{\beta^3}\right](3+5\beta^2) - \frac{6}{\beta^2} \nonumber\\
 - \beta^2 \ln \Rm - 2\beta^2 \ln\left(\frac{1+\beta^2}{1+\beta^2\sqrt{\Rm}}\right), \nonumber
\end{eqnarray}
\begin{eqnarray}
A_2^{(0)}(\beta) = \frac{2}{5}\left[2-\frac{\tan^{-1}\beta}{\beta^3}\right](9+5\beta^2) + \frac{9}{\beta^2} \nonumber\\
 - \beta^2 \ln \Rm - 2\beta^2 \ln\left(\frac{1+\beta^2}{1+\beta^2\sqrt{\Rm}}\right).
\label{eqA12}
\end{eqnarray}
The above \Eq{RKfit} is obtained by ignoring the effects of convection,
which corresponds to putting the parameter $a_* = 0$ in Eq.~(A21) of \cite{RK07}.
The superscripts $(0)$ indicate that these values are not affected
by the mean magnetic field, so they are the kinematic values,
hence $\Beqtwoz$ is the unquenched value of $\uu^2$ and
$\overline{\bb^2}^{(0)}$ is the contribution from the small-scale dynamo,
which is equal to zero as in this case there is no small-scale dynamo.
We compute $\overline{\bb^2}$ from \Eq{RKfit} and 
display it as a dashed line in \Fig{fig:tang_sat}.
We see that the analytical theory, which is valid for $\Rm\gg1$ and $\Rey\gg1$, 
is not far from our simulation data, which are at much smaller 
$\Rm$ ($\approx 20$) and $\Rey$ ($\approx 40$).
Our result generally confirms the theoretical prediction of 
the saturation of small-scale magnetic field at $\meanB_y > \Beq$.

Returning to the other runs in this regime (R2d--R2h), we notice that 
$\Rm$ is not much different in these runs, but the values of $D$
have increased systematically from Run~R2d to Run~R2h.
The small-scale field ($\tilde{\mean{b}}$) has also increased systematically in these runs. 
This clearly demonstrates that the tangling increases not only with $\Rm$,
but also with the large-scale field itself.
However, in all these runs, except in Run~R2d, we again observe the saturation of small-scale field
for $\meanB_y > \Beq$.

\subsection{Large $\Rm$ and vanishing $D$: Regime III}
\label{sec:regIII}

\begin{figure}[t]\centering
\includegraphics[width=\columnwidth]{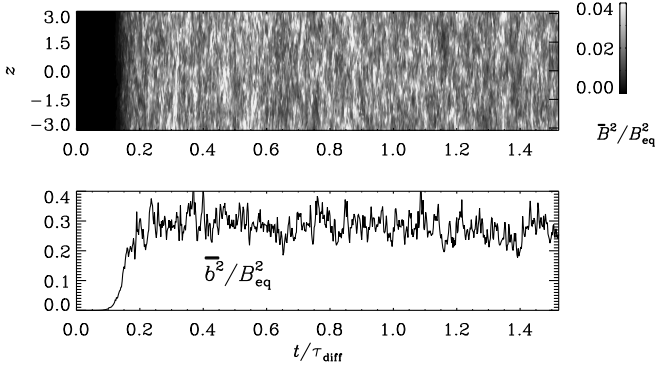}
\vspace{0.005in}
\caption{Results from Run~R3 in Regime~III: 
Butterfly diagram of the energy density of 
the large-scale magnetic field $\overline{\BB}^2$ (top) 
and time series of the small-scale field $\overline{\bb^2}$ at $z=0$ (bottom).
}\label{fig:R1}\end{figure}

We perform a simulation with
$\sigma = 0$ (non-helical), and $S = 0$.
Run~R3 in \Tab{tab:runs} represents this simulation.
When $C_\alpha = 0$ and $C_\Omega = 0$,
we should not expect a large-scale magnetic field to develop.
We have chosen such setup to exclude the effect
of the large-scale magnetic field in this simulation.
The top panel of \Fig{fig:R1} shows the energy of the large-scale field
$\overline{\BB}^2$.
We do not see any large-scale structures comparable to the system size.
In any case, the small-scale dynamo is expected to be excited as in
this case the magnetic Reynolds number $\Rm \sim 86$ is sufficiently
large for large values of the magnetic Prandtl number,
$\Pm = 5$, to excite the small-scale dynamo.
This is indeed confirmed by the present results.

In the bottom panel of \Fig{fig:R1}, we show the time series of the
energy density of the small-scale field.
Initially this field grows exponentially in time and then remains constant.
The characteristic length scale of this field is, as expected for a
small-scale dynamo, smaller than the forcing scale of the turbulence, as
can be seen from the magnetic energy spectrum in \Fig{fig:specR3R4a} (thick dashed curve).

%--------------- Regime IV -----------------------
\subsection{Large $\Rm$ and large $D$: Regime IV}
\label{sec:regIV}

\begin{figure}[t]\centering
\includegraphics[width=\columnwidth]{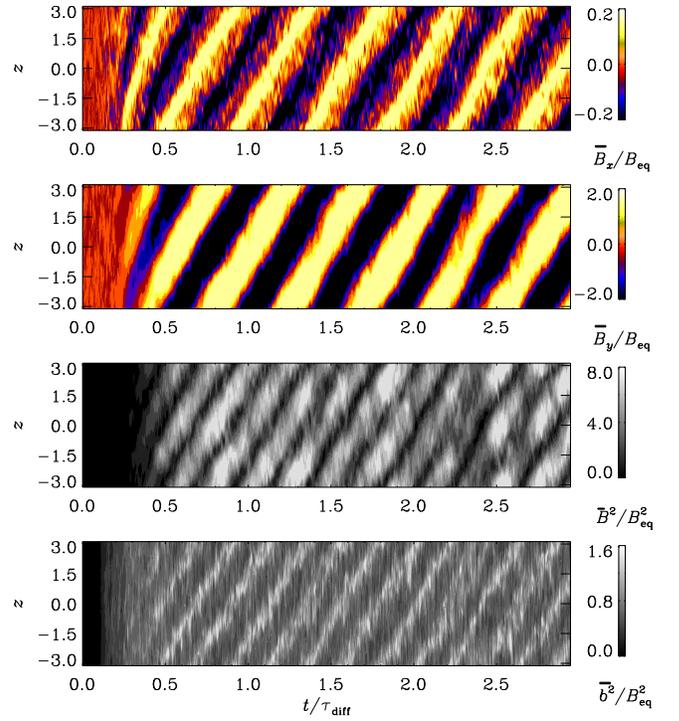}
\caption{Results from Run~R4a in Regime~IV: 
Butterfly diagrams of $\meanB_x$, $\meanB_y$, $\overline{\BB}^2$, 
and $\overline{\bb^2}$.
}\label{fig:R2a}\end{figure}

Our first two models in Regime~IV are similar to Run~R3, but with finite
shear and helicity such that the large-scale dynamo is excited.
We begin with $S=-0.05$ and $\sigma = 1$.
In \Tab{tab:runs}, we refer to this run as R4a.
The magnetic butterfly diagrams of various quantities are displayed in \Fig{fig:R2a}.
Again, we see dynamo waves of the large-scale field components
propagating in the positive $z$ direction.
The mean toroidal field ($\meanB_y$) is here about 10 times
stronger than the mean poloidal field ($\meanB_x$).

%-----Run: R4a-----

\begin{figure}[t]\centering
\includegraphics[width=\columnwidth]{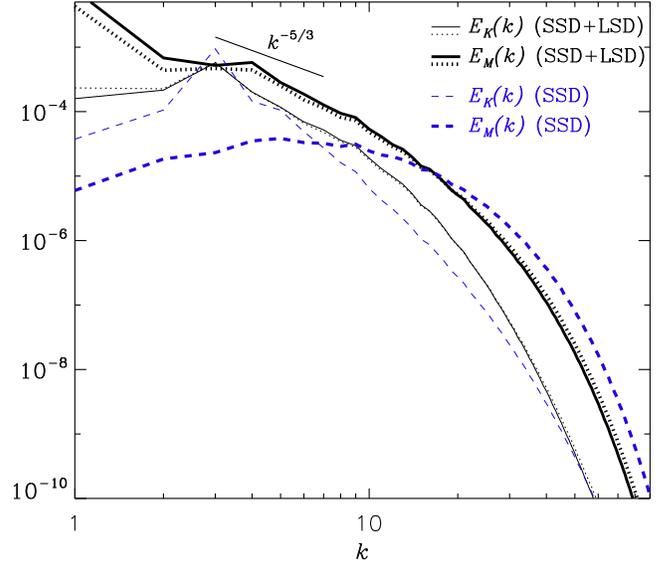}
\caption{
Evolutions of the energy spectra $E_K (k, t)$ (thin curves) and $E_M (k, t)$ (thick) 
for Run~R3$^\prime$ (SSD, dashed curves) and Run~R4a$^\prime$ (SSD+LSD, solid and dotted
lines correspond to magnetic maxima and minima, respectively).
All spectra are temporally averaged excluding the initial growing phase. 
}\label{fig:specR3R4a}\end{figure}

\begin{figure}[t]\centering
\includegraphics[width=\columnwidth]{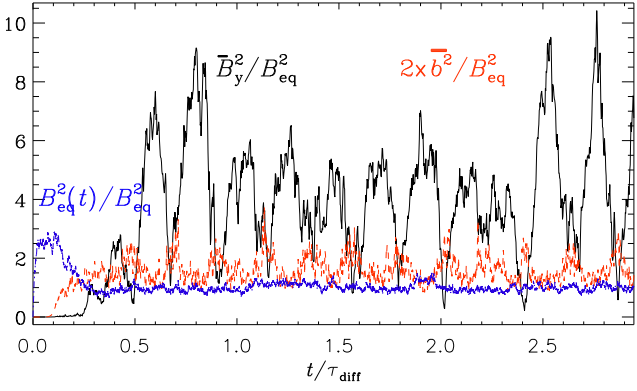}
\caption{Results from Run~R4a: Time series of $\overline{B}_y^2$ 
(solid black line), $\overline{\bb^2}$ (red dashed), and $\Beq^2(t)$ (blue dash-dotted). 
Note that $\overline{\bb^2}/\Beq^2$ is rescaled by a factor of two.
}\label{fig:R2ats}\end{figure}
\begin{figure}[t]\centering
\includegraphics[width=\columnwidth]{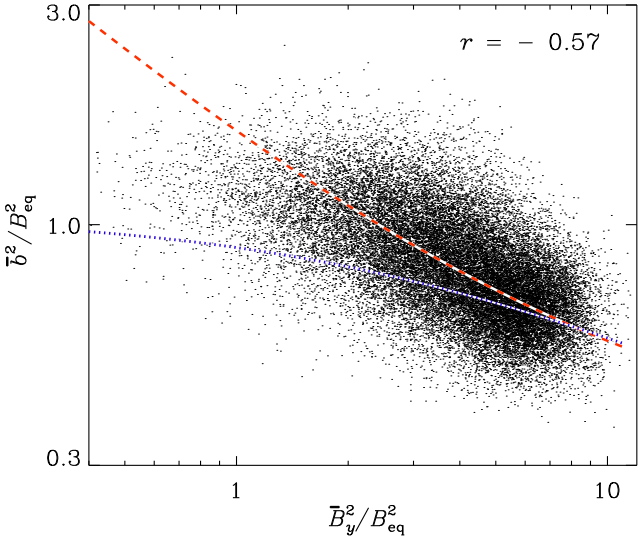}
\caption{Results from Run~R4a: A scatter plot between $\meanB_y^2/\Beq^2$, 
and $\overline{\bb^2}/\Beq^2$ at different values of $z$ coordinate.
The red dashed line shows the $\overline{\bb^2} \propto \meanB_y^{-1.4}$ dependence,
whereas blue dotted line represents \Eq{RKfit}.
The data points for initial growth phase ($t/\tau_{\rm diff} < 0.6$)
are not used for this plot.
The linear correlation coefficient $r$ is marked on the plot.
}\label{fig:R2asct}\end{figure}

In \Fig{fig:specR3R4a}, we show for Run~R4a$^\prime$ (a high resolution
version of Run~R4a) kinetic and magnetic energy
spectra, $E_K (k, t)$ (thin solid and dotted) and $E_M (k, t)$ (thick
solid and dotted), during magnetic maximum and minimum, respectively.
For comparison we also show the spectra for a small-scale dynamo from 
Run~R3$^\prime$ (a high resolution version of Run~R3).
For both dynamos, $E_K (k, t)$ has a peak at the forcing wavenumber 
which is around 3 in these cases, followed by a small
inertial range, that is slightly steeper than $k^{-5/3}$,
and finally the energy dissipation at small scales.
However, $E_M (k, t)$ shows different behaviors in the two cases.
For the LSD there is an intermediate peak slightly below the forcing scale ($k\approx4$)
and then there is a maximum at the largest scale ($k=1$), as expected for the large-scale dynamo.
During magnetic maximum, $E_M (k, t)$ is larger at large scales
(compare thick solid and dotted curves), 
but it is slightly smaller at small scales ($k > 20$). 
However, during magnetic maximum $E_K (k, t)$ is slightly smaller at larger scale 
($k <3 $; compare thin solid and dotted curves).

Owing to the cyclic variation of the large-scale field we see a weak 
modulation of the energy of the small-scale field, 
as seen in the lower two panels of \Fig{fig:R2a}. 
In fact, $\overline{\bb^2}$ shows a weak anticorrelation with $\overline{\BB^2}$.
This is evident from \Fig{fig:R2ats}, where time series of $\meanB_y^2$ 
and $\overline{\bb^2}$ at $z=0$ are shown.
From \Fig{fig:R2ats} we note that $\overline{\bb^2}$ reaches
significant levels much faster than the large-scale energy.
This implies that initially $\overline{\bb^2}$ is primarily a result of the small-scale dynamo,
which grows faster than the large-scale dynamo. 
Later, with the development of the large-scale field, $\overline{\bb^2}$ increases further. 
Eventually, in this run, the saturated value of $\overline{\bb^2}$ is much larger than
that in Run~R3 (which is basically the same as in this run, but only the small-scale dynamo is excited
because of taking $\sigma=0$, and $S=0$; see \Tab{tab:runs}).

Considering only the nonlinear stage ($\meanB_y \ge \Beq$), we produce
in \Fig{fig:R2asct} a scatter
plot between $\meanB_y^2$ and $\overline{\bb^2}$ for all grid points along 
the $z$ direction.
This shows an anticorrelation obeying $\overline{\bb^2} \propto \meanB_y^{\,-1.4}$;
see the red dashed line.
The linear correlation coefficient between $\meanB_y^2$ and $\overline{\bb^2}$  is $-0.57$.
It is not straightforward to understand the origin of such variations:
the small-scale magnetic field developed in this simulation can be read as
$\overline{\bb^2} = \overline{\bb^2}_{\rm SSD} + \overline{\bb^2}_{\rm tang}$,
where $\overline{\bb^2}_{\rm SSD}$ results purely from the small-scale dynamo, while
$\overline{\bb^2}_{\rm tang}$ results from the tangling of the large-scale field.
Unfortunately, separating these two quantities from a given simulation is not an easy task 
because both quantities are affected by the large-scale field $\meanBB$. 
In the initial phase, when the large-scale field is not yet fully developed,
the small-scale field is mainly a consequence of the small-scale dynamo. 
Thus, in this phase we have $\overline{\bb^2} \approx \overline{\bb^2}_{\rm SSD}$.
Then, with increasing $\meanBB^2$, the small-scale field is increasing
further because of the contribution from the tangling of the large-scale field. 
However, when $\meanBB^2$ exceeds $\Beq^2$, $\overline{\bb^2}_{\rm tang}$ tends to saturate.
(Such behavior was seen in earlier simulations of \Sec{sec:regII}
showing that, when the small-scale dynamo was not operating,
the small-scale field generated due to tangling of the large-scale field increased
with the mean-field first and then saturated; see \Fig{fig:tang_sat}.)
Therefore, in the nonlinear phase of the present simulation, the anticorrelation 
results from the fact that the 
small-scale dynamo is suppressed by the large-scale magnetic field
through the Lorentz force acting on the flow. 
However, when we look at the energy of the turbulence, $\Beq^2(t)$ ($=\uu^2$), as shown in \Fig{fig:R2ats},
we see no anticorrelation between the flow and the large-scale field.
This is possibly because our turbulence is continuously forced by a forcing
function which could break the anticorrelation between $\meanB_y^2$ and $\uu^2$.

Keeping in mind that the theory of \citet{RK07} was developed for magnetic fluctuations
generated due to tangling and a small-scale dynamo represented by
a constant factor $\overline{\bb^2}^{(0)}$, which is not the case here, we now try to fit our data by choosing
different values of $p$ and $q$ in \Eq{RKfit}.
We find that, when $\overline{\bb^2}^{(0)} = 1.02 \Beqtwoz$, $c_1=4.0$, and $c_2=0.5$, 
there is moderate agreement, but only for $2<\meanB_y^2/\Beq^2<10$; see blue dotted line in \Fig{fig:R2asct}.

\begin{figure}[t]\centering
\includegraphics[width=\columnwidth]{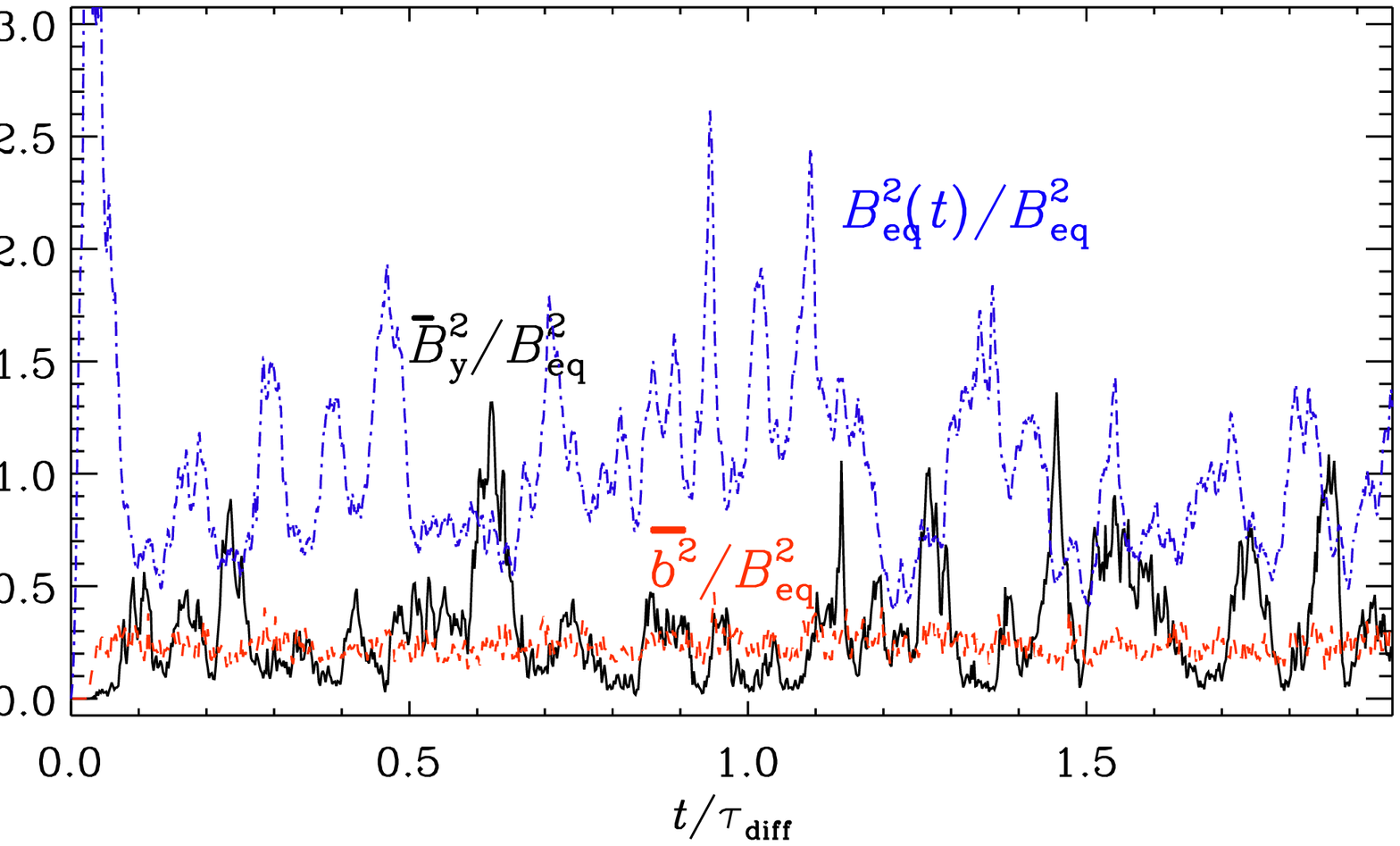}
\caption{Same as \Fig{fig:R2ats} but results from R4g.
}\label{fig:R2bts}\end{figure}
\begin{figure}[t]\centering
\includegraphics[width=\columnwidth]{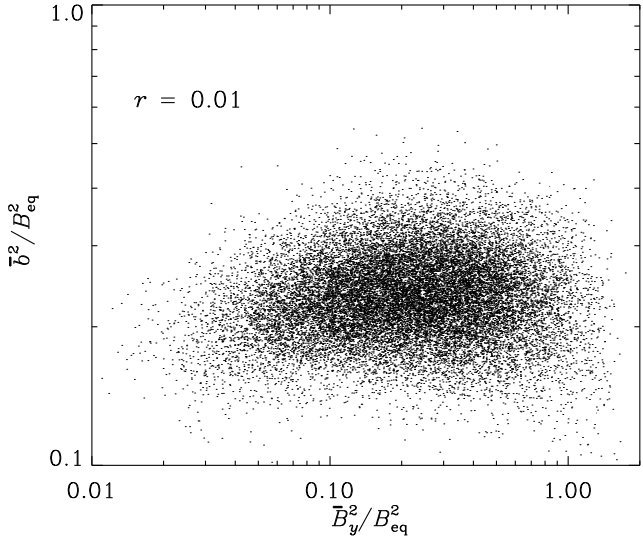}
\caption{Same as \Fig{fig:R2asct} but results from Run~R4g.
The data points for initial growth phase ($t/\tau_{\rm diff} < 0.3$)
are not used for this plot.
}\label{fig:R2bsct}\end{figure}

We have performed a few more simulations in this regime (see Runs R4b--R4h in \Tab{tab:runs}) 
at different values of $\Rm$, $\Pm$, or $D$.
We observe that, as long as the large-scale field is above $\Beq$,
the small-scale field is anticorrelated with the large-scale field.
This is the case for Runs R4a--R4f.
However for Runs R4g and R4h we see that the large-scale dynamo is much
weaker (compare the $D$ values) and consequently the large-scale field is
much weaker than $\Beq$ so that it cannot affect the small-scale field production significantly.
Therefore, we do not see any anticorrelation
between small- and large-scale fields in this case.
In \Fig{fig:R2bts} we show the time series of $\meanB_y^2$ and $\overline{\bb^2}$
and in \Fig{fig:R2bsct} the corresponding scatter plot for Run R4g.
Here we do not see any anticorrelation between the small-scale and the large-scale fields.
Another feature to note is that in the earlier Runs~R4a--R4f the small-scale field 
$\tilde{\mean b}$ was more than $82\%$ of $\Beq$ (see \Tab{tab:runs}), 
whereas in Runs R4g--R4h the small-scale field is only about $48\%$ of $\Beq$. 
We would expect the opposite because the small-scale dynamo should be much stronger 
as $\Rm$ is larger in the latter two runs. Therefore, larger values of $\tilde{\mean b}$ 
in Runs R4a--R4f compared to Runs R4g--R4h imply that a significant contribution 
of small-scale field in Runs R4a--R4f is coming from tangling.

On this occasion we mention that, using the same setup as here, \cite{HRB11} 
have demonstrated that, after some initial saturation phase, the $\alpha\Omega$ 
dynamo mode is ``fratricided'' by a separate $\alpha^2$ mode and eventually the 
initial $\alpha\Omega$ model changes to an $\alpha^2$ type dynamo,
which is sustained by the helical turbulence alone.
Some of the simulations, particularly
at larger $\Rm$ and large $D$ (Runs~R4a--R4f),
display such behavior. 
Therefore, analyzing these simulations for longer durations is problematic.
In this connection it maybe useful to note that the
typical wall clock time per timestep per mesh point for simulations with $144^3$ 
meshpoints using 48 processors is about $0.07~\mu$s. Therefore, for example,
to produce results for Run R4a (presented in \Fig{fig:R2a}) it took about 24~hours,
while for R4g (\Fig{fig:R2bts}) it took 43~hours.
As examples, for Runs R2a and R4g, $\tau_{\rm diff}/\Delta t \approx$ 2400 and 410000, respectively.

\section{Summary and conclusions}

We have performed a set of simulations in different parameter regimes
of Reynolds and dynamo numbers to explore the effect of the cyclic variation
of the dynamo-generated large-scale magnetic field on the small-scale field.
Our ultimate motivation is to shed some light on the solar cycle
dependence of the small-scale magnetic field at the solar surface.
In the Sun, both global and local dynamos are expected to operate at
the same time and can in principle occur even in the same location.
To study this problem in detail, it is necessary to develop a model
in which both dynamos are operating.
We have therefore considered a simple turbulent dynamo in which
the fluid is helically forced and large-scale linear shear is imposed.
This simple setup captures the essential mechanism of an $\alpha \Omega$ dynamo
believed to be operating in the Sun.
At the same time, the local dynamo can be excited in this setup when $\Rm$ exceeds
a critical value.
The main advantage of considering this type of simple setup, 
rather than considering more sophisticated global convection simulations in 
spherical geometry (Hotta et al., private communication)
are an affordable computational expense and that we can easily change the
large-scale dynamo parameters to explore different  
regimes of these two dynamos.

In Regime~III, when the large-scale dynamo number is small but the Reynolds number is 
large, we observe only the small-scale dynamo in which the generated small-scale field 
does not have any temporal variation. This situation does not apply to the Sun
as large-scale magnetic field is also generating by the global dynamo.
In Regime~IV (\Sec{sec:regIV}), when both the Reynolds and dynamo numbers are 
large\footnote{In the upper layer of the Sun, which is our region of
interest, we have $\Rm \sim 10^8$---much larger than what we consider. However,
based on stellar observations, we may imagine that in the Sun, $D$ is not much
larger than the critical dynamo number needed to excite the large-scale
dynamo; see Section~5 of \citet{KKB15}.}, we observe both
dynamos---the large-scale dynamo producing an oscillatory large-scale magnetic field
and the local dynamo producing a small-scale field.
This situation is relevant to the Sun.
In our simulations, we observe that the small-scale field is 
almost uncorrelated with the dynamo cycle as long as the large-scale field 
is smaller than the equipartition value of turbulence.
However, when the large-scale field is stronger than the equipartition value, we observe that 
the generation of the small-scale field is quenched by the large-scale field---producing a weak anticorrelation 
between small- and large-scale fields.
This reciprocal correlation between the small-scale field and the global dynamo cycle is 
in agreement with observations of the small-scale magnetic field at the solar surface
\citep[e.g.,][]{HST03,Jin11,JW12,JW15b,FR15}.
In particular, \citet{Jin11} have analyzed separately the small-scale elements 
based on their magnetic flux contents and the number of magnetic elements
and found that their cyclic variations
for magnetic elements with fluxes
in the range $\sim(3$--$30) \times10^{18}\Mx$ and
$\sim(4$--$40) \times 10^{19}\Mx$ show anticorrelation 
and correlation with sunspots, respectively.
Further analysis by \citet{JW12} has revealed that this anticorrelated component 
shows a solar cycle variation different from the sunspot variation
in time and latitude,
whereas the correlated component shows similarities with the sunspot butterfly diagram.
Based on our simulation results in Regime~IV, and
the observational results, we conclude that the small-scale magnetic
field observed at the solar surface \citep[particularly, the magnetic elements
with fluxes $\le 3\times 10^{19}\Mx$ in the analysis of][]{Jin11}
are a consequence of both the small-scale dynamo generated
locally in the solar convection zone and the shredding and tangling of the large-scale
magnetic field generated by the global dynamo.
On the other hand, the magnetic field with a flux above
$4\times 10^{19}\Mx$, but smaller than the active region flux
in the analysis of \citet{Jin11},
are possibly the result of shredding and tangling of the large-scale
magnetic field generated by the global dynamo only.
Our simulations in Regime~II (\Sec{sec:regII}), when the dynamo number is
large but the Reynolds number is small, illustrate this situation because in
this case the local dynamo is not possible and the small-scale field is
still generated by the tangling of the large-scale magnetic field.

We reiterate that we observe a weak anti-correlation between small-scale field
and large-scale magnetic cycle only when the large-scale field is much stronger
than the equipartition value. Based on our knowledge, we can say that the large-scale field
in the solar convection zone may be larger than the equipartition value,
although probably not by much.
Therefore we expect that the small-scale magnetic field in the Sun is largely unaffected
by the global large-scale cycle, but not completely.
The lack or weakness of a variation of the 
small-scale magnetic field observed in our simple turbulent dynamo simulations
certainly support the idea that both the small-scale dynamo and the tangling of large-scale field are responsible for the observed 
small-scale magnetic field in the Sun. In fact, in certain regimes our results are in 
agreement with observations of the small-scale magnetic field.
However, future studies using more realistic solar dynamo models will be useful for 
more quantitative comparisons.

\begin{acknowledgements}
BBK is grateful to Sami K.\ Solanki for the hospitality at Max Planck Institute 
for Solar System Research when the motivation of this project was conceived
during a visit back in 2011.
We thank the referee for his/her critical comments that have improved 
the presentation. 
We also thank Igor Rogachevskii for the discussion in comparing with analytical theory.
Finally, we acknowledge the allocation of computing resources provided by the
Swedish National Allocations Committee at the Center for
Parallel Computers at the Royal Institute of Technology in
Stockholm and the National Supercomputer Center in Ume\aa.
This work was supported in part by
the Swedish Research Council grants No.\ 621-2011-5076 and 2012-5797,
as well as the Research Council of Norway under the FRINATEK grant 231444.
\end{acknowledgements}

\bibliographystyle{apj}
\bibliography{paper}
\end{document}

%% file: aa.tex
\newcommand{\EQ}{\begin{equation}}
\newcommand{\EN}{\end{equation}}
\newcommand{\EQA}{\begin{eqnarray}}
\newcommand{\ENA}{\end{eqnarray}}

\newcommand{\Eq}[1]{Equation~(\ref{#1})}

\newcommand{\Sec}[1]{Section~\ref{#1}}

\newcommand{\Fig}[1]{Figure~\ref{#1}}

\newcommand{\Tab}[1]{Table~\ref{#1}}

\newcommand{\bra}[1]{\langle #1\rangle}

\newcommand{\mean}[1]{\overline #1}

{}
{}
{}

{}
{}
{}
{}
{}
{}
{}
{}
{}
\newcommand{\meanBB}{\overline{\mbox{\boldmath $B$}}{}}{}
{}
{}
{}
{}
{}
{}
{}
{}
{}
{}
{}

{}
{}
{}

\newcommand{\meanB}{\overline{B}}

{}

{}
{}

\newcommand{\eee}{\hat{\mbox{\boldmath $e$}} {}}

\newcommand{\kk}{\bm{k}}

\newcommand{\xx}{\bm{x}}

\newcommand{\bb}{\bm{b}}
\newcommand{\BB}{\bm{B}}

\newcommand{\uu}{\mbox{\boldmath $u$} {}}
\newcommand{\UU}{\mbox{\boldmath $U$} {}}

\newcommand{\JJ}{\mbox{\boldmath $J$} {}}

\newcommand{\AAA}{\mbox{\boldmath $A$} {}}

\newcommand{\ff}{\mbox{\boldmath $f$} {}}

\newcommand{\nab}{\mbox{\boldmath $\nabla$} {}}
\newcommand{\RRRR}{\mbox{\boldmath ${\sf R}$} {}}
\newcommand{\SSSS}{\mbox{\boldmath ${\sf S}$} {}}

\newcommand{\ii}{{\rm i}}

\newcommand{\DIV}{\bm{\nabla} \cdot }

\def\Pm{P_{\rm m}}

\def\Rm{R_{\rm m}}

\def\Rey{\mbox{\rm Re}}

\def\cs{c_{\rm s}}

\def\kf{k_{\rm f}}
\def\urms{u_{\rm rms}}

\def\etatz{\eta_{\rm t0}}
\def\etaTz{\eta_{\rm T0}}

\def\Beq{B_{\rm eq}}

\def\Beqtwoz{B_{\rm eq}^{2\,(0)}}

\def\onethird{{\textstyle{1\over3}}}

\newcommand{\Mx}{\,{\rm Mx}}

\def\onethird{{\textstyle{1\over3}}}
\def\onehalf{{\textstyle{1\over2}}}

\def\onethird{{\textstyle{1\over3}}}

\newcommand{\BoldVec}[1]{\mathchoice%
  {\mbox{\boldmath $\displaystyle     #1$}}%
  {\mbox{\boldmath $\textstyle        #1$}}%
  {\mbox{\boldmath $\scriptstyle      #1$}}%
  {\mbox{\boldmath $\scriptscriptstyle#1$}}%
}